\documentclass{article}

\pdfoutput=1

\usepackage[preprint]{neurips_2026}

\usepackage[utf8]{inputenc} 
\usepackage[T1]{fontenc}    
\usepackage{hyperref}       
\usepackage{url}            
\usepackage{booktabs}       
\usepackage{amsfonts}       
\usepackage{nicefrac}       
\usepackage{microtype}      
\usepackage{xcolor}         
\usepackage{amsmath}
\usepackage{xspace}
\usepackage{amsmath}
\usepackage{arydshln}
\usepackage{enumitem}
\usepackage{wrapfig} 
\usepackage[table]{xcolor}
\setlist[itemize]{leftmargin=*}
\setlist[enumerate]{leftmargin=*}
\usepackage{mathtools}
\usepackage{hyperref}
\usepackage{tcolorbox}

\title{LLM Agents Enable User-Governed Personalization Beyond Platform Boundaries}

\author{%
Jiacheng Lin$^{1}$\thanks{Main contributors. Authors are listed in alphabetical order by last name.}\quad Kun Qian$^{2,*}$ \quad
Arvind Srinivasan$^{3,*}$ \quad
Tian Wang$^{4,*}$ \quad 
Fang Han$^{3}$ \quad \\
\textbf{Changran Hu}$^{5}$ \quad
\textbf{Junze Liu}$^{8}$ \quad
\textbf{Ziyi Wang}$^{6}$ \quad
\textbf{Hanwen Xu}$^{7}$ \quad
\textbf{Mengmeng Xue}$^{6}$ \quad 
\textbf{Shuo Yang}$^{2}$ \quad \\
\textbf{Hansi Zeng}$^{10}$ \quad
\textbf{Simon Sinong Zhan}$^{11}$ \quad
\textbf{Kai Zhong}$^{2}$ \quad
\textbf{Weiqi Zhang}$^{9}$ \quad
\textbf{Dakuo Wang}$^{6}$ \quad \\
\textbf{Tianhao Wang}$^{12}$ \quad
\textbf{Zhiyuan Li}$^{13}$ 
\\
\\
$^{1}$ University of Illinois Urbana Champaign \hspace{1em}
$^{2}$ University of Texas at Austin \hspace{1em} \\
$^{3}$ Carnegie Mellon University \hspace{1em}
$^{4}$ New York University \\
$^{5}$ University of California, Berkeley \hspace{1em}
$^{6}$ Northeastern University \hspace{1em}
$^{7}$ University of Washington \\
$^{8}$ University of California, Irvine \hspace{1em}
$^{9}$ University of New Brunswick \\
$^{10}$ University of Massachusetts at Amherst \\
$^{11}$ Northwestern University \hspace{1em}
$^{12}$ University of California, San Diego \\
$^{13}$ Toyota Technological Institute at Chicago
}

\begin{document}

\maketitle

\begin{abstract}
Personalization today is fundamentally platform-centric: services build user representations from the behavioral fragments they observe. Yet no platform can construct a complete picture of the user, as competitive incentives, legal constraints, user privacy concerns, and epistemic limits create persistent data barriers. This paper argues for a shift from platform-centric personalization to user-governed personalization, where only the user can integrate fragmented contexts across platforms and the offline world. The key asymmetry lies in data access: only users can aggregate their own cross-platform and offline information. Large language model (LLM) agents make such integration practically feasible for the first time by enabling reasoning over heterogeneous personal data and transforming users' cross-context information into actionable personalization capabilities. We provide proof-of-concept evidence that users equipped with cross-platform data exports and an off-the-shelf LLM agent can outperform single-platform personalization baselines. We conclude by outlining a research agenda for building scalable user-governed personalization systems.
\end{abstract}

\section{Introduction}

Personalization, broadly defined, is the process of tailoring a system's content and services to individual users based on knowledge of their preferences, needs, and behavior \citep{adomavicius2005personalization, fan2006personalization, adomavicius2005toward, zhang2025personalization}. By reducing information overload and surfacing content that matches individual interests, effective personalization shapes how people navigate the increasingly complex digital world \citep{xiao2007commerce,tam2006understanding,zhang2025personalization}. Major platform companies have invested heavily in personalization, such as Netflix and TikTok. More recently, large language models (LLMs) have enabled platforms to reason over user data across multiple services in ways that were previously infeasible. Apple launched its AI strategy under the banner of ``personal intelligence'' that reason across messages, emails, photos, and calendars \citep{apple2024intelligence}, and Google allows users to connect their personal data across Gmail, Photos, Search, and Maps to Gemini \citep{pichai2025googleio}. These efforts reflect a shared direction: pursuing more comprehensive understanding of users to enable better personalization \citep{adomavicius2005personalization}.

Although platforms are pursuing increasingly comprehensive user understanding, each platform can observe only the slice of a user’s life that occurs \emph{within its own ecosystem}. Yet users’ preferences, needs, and life contexts are not confined to any single platform. They are distributed across many digital services and, crucially, across the offline world. For example, Amazon may know what a user buys, but it does not necessarily know where the user goes, what they search for outside Amazon, whom they interact with, or how their daily routines and life circumstances are changing. As a result, no single platform, regardless of the sophistication of its algorithms, can construct a complete picture of the user. \textit{This limitation matters because prior research has long emphasized that personalization depends on comprehensive and holistic user modeling} \citep{adomavicius2005personalization,kobsa1994user,berkovsky2008mediation,cantador2015cross,purificato2024user}.  Moreover, this fragmentation cannot be resolved through platform data sharing: competitive incentives make user data a proprietary moat that no platform will voluntarily open, no platform can observe a user's offline context, life events, or the motivations behind their choices, and privacy concerns further limit the scope of data that users are willing to share across platforms. 

Against this backdrop, the only entity that naturally spans all of these fragmented contexts is the user. Unlike any single platform, the user directly experiences the full continuity of their own life: their activities across services, their offline routines, their shifting goals, and the private motivations that underlie observable behavior. In principle, this makes the user the most complete and authoritative source of information for personalization. Yet historically, this potential has remained largely unrealized. Although users generate and implicitly hold access to rich, cross-contextual data about themselves, they have lacked the tools to aggregate, interpret, and operationalize this information at scale. Personal data has therefore been abundant but inert—captured in silos, difficult to integrate, and rarely leveraged directly by users to shape their own digital experiences.

\begin{figure}[t]
    \centering
    \includegraphics[width=\linewidth]{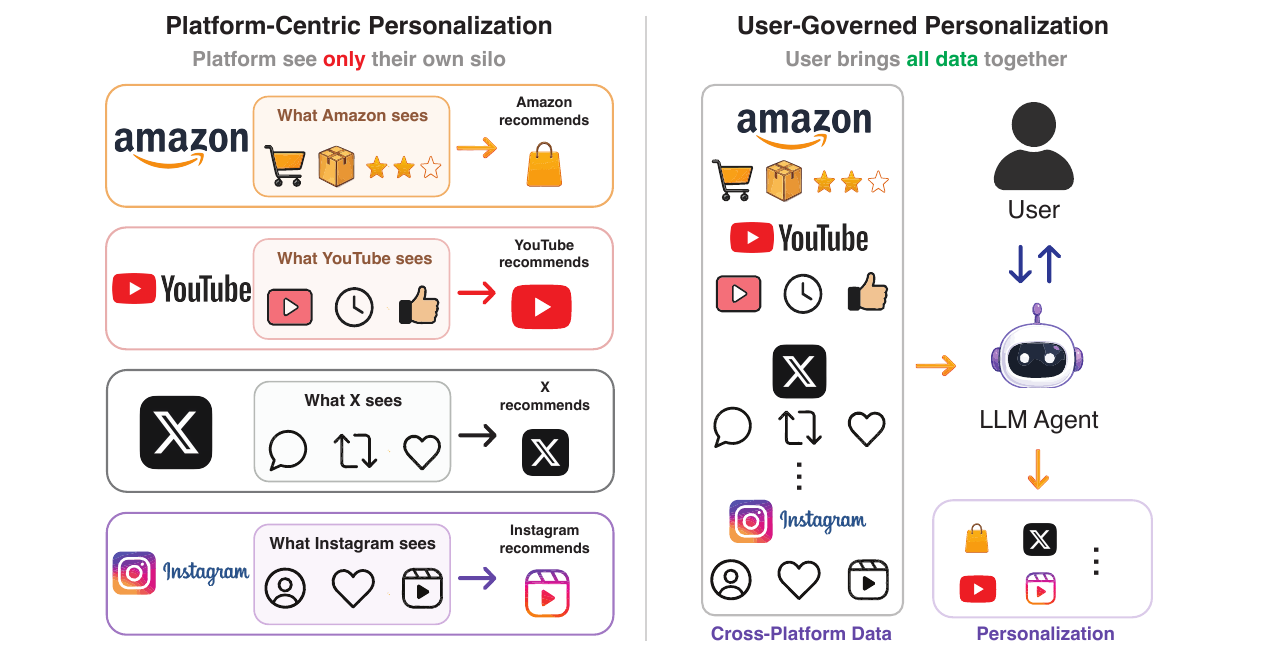}
    \caption{\textbf{Platform-centric vs. user-governed personalization.} Left: each platform observes only the behavioral data generated within its own service and personalizes in isolation. Right: the user aggregates data from multiple platforms and delegates personalization to
   an LLM agent, which reasons over the combined cross-platform context to produce recommendations. }
    \label{fig:placeholder}
    \vspace{-1em}
\end{figure}

LLM agents \citep{anthropicClaudeCode,openai2025codex} fundamentally change this equation. By enabling reasoning over heterogeneous data, cross-service interaction, and user-side action, LLM agents turn the user’s previously underutilized informational advantage into an actionable capability. Rather than relying on platforms to infer preferences from partial traces, users can now orchestrate personalization across a more holistic understanding of their lives. This shift moves personalization from a platform-centric optimization problem to a user-governed process, where users actively integrate data across boundaries that platforms cannot cross. Consequently, comprehensive personalization is achievable through the user, and is now starting to become practically attainable. \textbf{In this paper, we argue for a shift from platform-centric personalization to user-governed personalization, where only the user can integrate fragmented contexts, and where LLM agents make such integration both scalable and actionable for the first time}\footnote{We do not argue that platforms become irrelevant. Under current architectures, platforms continue to serve as content providers and matching layers (e.g., candidate retrieval, inventory management); user-governed personalization operates on top of this infrastructure, with the user-side LLM agent determining the final ranking and selection based on a more complete picture of the user.}.

We first show that platform personalization faces a data barrier: competitive incentives, legal constraints, and epistemic limits ensure that no single platform can access a complete picture of any user (Section \ref{sec:data_barrier}). We then argue that the user is the only entity capable of integrating data across these boundaries, and that LLM agents transform users' previously underutilized cross-platform data advantage into an actionable capability for the first time; we support this argument with empirical evidence showing that users equipped with cross-platform data exports and an off-the-shelf LLM can outperform single-platform baselines (Section \ref{sec:proposal}). In Section \ref{sec:alternative}, we consider objections. We conclude with open problems and future directions for the machine learning community, platform developers, and policymakers (Section \ref{sec:agenda}).

\vspace{-0.5em}
\section{Data Barrier in Platform Personalization}
\vspace{-0.5em}
\label{sec:data_barrier}
\subsection{How Platforms Personalize}
Personalization is central to how digital platforms create value. Over the past three decades, platforms have developed a wide range of methods to tailor content, products, and services to individual users.

\textbf{Collaborative filtering and matrix factorization.} Alongside content-based methods that match user profiles to item attributes, collaborative filtering (CF) became the dominant paradigm in platform personalization since the early 1990s \citep{goldberg1992using, resnick1994grouplens}. Rather than attempting to build a deep model of each individual user, CF infers preferences from behavioral patterns across the entire user population: users who agreed in the past are likely to agree in the future. Item-based variants scaled this approach to millions of products at Amazon \citep{linden2003amazon}, while matrix factorization methods, catalyzed by the Netflix Prize competition, learned compact latent representations from large-scale user-item interaction matrices \citep{koren2009matrix}. The core logic of CF remains influential today: leverage population-level patterns to compensate for sparse individual user profiles.

\textbf{Deep learning at scale.} Beginning in the mid-2010s, deep learning transformed production recommendation systems. YouTube deployed a two-stage architecture combining deep candidate generation with deep ranking models \citep{covington2016deep}.  Google introduced Wide\&Deep learning, jointly training wide linear models for memorization and deep neural networks for generalization \citep{cheng2016wide}. Meta developed the Deep Learning Recommendation Model (DLRM), which processes both dense user features and sparse categorical features through embedding tables and interaction layers \citep{naumov2019deep}. These systems operate on billions of user interactions and are continuously retrained to capture shifting behavioral patterns.

\textbf{LLM-based personalization.} Most recently, platforms have begun integrating user data across their own services to build more comprehensive user models \citep{apple2024intelligence,pichai2025googleio}. In parallel, large language models have been applied to recommendation tasks, enabling reasoning over user histories in natural language rather than through learned embeddings alone \citep{wu2024survey,lin2025recr}. These developments represent the current frontier of platform personalization: platforms are actively pursuing more holistic user understanding by combining data across their own product ecosystems.

Despite this steady methodological evolution, from collaborative filtering through deep learning to LLM-based approaches, \textit{all platform personalization operates on behavioral data collected within the platform's own boundaries}. Even cross-service integration efforts by Apple and Google remain confined to each company's own ecosystem. In the next subsection, we argue that this data constraint is not a temporary technical gap but a structural condition that no single platform can overcome.

\vspace{-0.5em}
\subsection{Why the Data Barrier Persists}
\vspace{-0.5em}
The methods described in the previous subsection can only personalize based on data that the platform itself observes. Because each platform sees only a fragment of any user's behavior, these methods compensate by learning from patterns across the entire user population. This population-level approach is effective for capturing broad preferences, but it reaches a fundamental limit when personalization requires fine-grained individual-level understanding \citep{hu2008collaborative,kleinberg2024challenge}. This limitation naturally raises the question of whether platforms can overcome it by acquiring more complete data about each user. In practice, however, such a path is constrained by persistent barriers that prevent any single platform from achieving a truly comprehensive view.

\textbf{Competitive barriers.} User data is among the most valuable competitive assets a platform possesses. A self-reinforcing cycle emerges: more users generate more behavioral data, which improves models trained on such data, which is utilized to attract more users \citep{gregory2021role,parker2016platform}. In this environment, cross-platform data sharing is a dominated strategy. Formal analysis of data sharing as an iterated game shows that cooperation requires external enforcement mechanisms that do not currently exist; data-driven markets tend toward concentration precisely because leading platforms have no incentive to share the data that sustains their advantage \citep{prufer2021competing}. Even when regulators mandate interoperability or data portability, platforms have both the incentive and the resources to comply minimally. Facebook's progressive restriction of its APIs from open access to tightly controlled tiers illustrates this dynamic: data access was curtailed as the competitive value of user data became apparent \citep{van2022api}.

\textbf{Regulatory reinforcement.} Regulation further constrains platforms' ability to combine user data. The Digital Markets Act (DMA) prohibits designated gatekeepers from combining personal data across their core platform services without specific user consent (Article 5(2)) \citep{eu2022dma,geradin2022interplay}, and enforcement has been strict: in April 2025, the European Commission fined Meta EUR 200 million for circumventing this requirement through a ``consent or pay'' model \citep{ec2025applemeta_dma}. Similar restrictions are emerging globally, including the California Privacy Rights Act's cross-context advertising opt-out provisions \citep{california2018ccpa1798}. While user consent can in principle bypass these constraints, the practical effect is to increase the cost and friction of building comprehensive user profiles, reinforcing the competitive barriers described above.

\textbf{User privacy concerns.} Users are fundamentally reluctant to have their personal data shared beyond the service where it was generated. Surveys consistently find that a majority of internet users are concerned about how companies collect and use their data \citep{auxier2019americans}. Any scheme that requires user data to flow to third parties, whether to other platforms or to centralized intermediaries, faces this fundamental demand-side resistance.

\textbf{Epistemic barriers.} All platform data, by definition, consists of behavioral traces recorded within digital services. These traces systematically miss critical categories of information that matter for personalization. First, cross-domain preference correlations are distributed across platforms and invisible to any single one: a user's music listening on Spotify, product searches on Amazon, and travel planning on Google each reveal different facets of the same underlying interests, but no platform observes the full pattern \citep{zhu2021cross,cantador2015cross}. Second, platforms observe what users do but not why. The same purchase can reflect fundamentally different motivations, and without access to the user's reasoning, platforms cannot distinguish between them; research on implicit feedback has formally characterized this limitation, showing that observed actions are inherently ambiguous and serve as noisy confidence signals rather than direct preference indicators \citep{hu2008collaborative}. Third, and most fundamentally, offline life context is entirely absent from platform logs. Life events such as relocating, changing careers, or starting a family shift preferences across every domain simultaneously, yet no digital service records these events. Context-aware recommendation research has long identified such factors as "unobservable context" that lies beyond the reach of any data collection mechanism \citep{adomavicius2010context}. Unlike competitive incentives and regulatory constraints, the epistemic barrier does not depend on institutional arrangements and cannot be addressed through policy. Even a hypothetical platform with access to every other platform's data would remain unable to observe users' offline lives, private motivations, and the reasoning behind their choices.

These forces do not operate independently; they reinforce one another. Competitive incentives prevent data from flowing between platforms. Users' own privacy concerns discourage sharing personal data. Regulation increasingly constrains data combination. Epistemic limits ensure that even unrestricted data pooling would leave the picture incomplete. The data barrier in platform personalization is therefore persistent: it cannot be overcome by more powerful algorithms, larger models, or incremental policy reforms. Any path to personalization beyond platform boundaries requires a fundamentally different approach to data access.

\vspace{-0.5em}
\section{Our Proposal: User-Governed Personalization}
\label{sec:proposal}
\vspace{-0.5em}

\subsection{Why Only Users Can Bridge the Data Barrier}
\vspace{-0.5em}
Section \ref{sec:data_barrier} established that platform personalization faces a persistent data barrier. This section argues that the user is the only entity capable of bridging this barrier, and that LLM agents make this capability practically actionable for the first time.

\textbf{The user as the unique integration point.} The user is the only entity that exists simultaneously across all platforms and the offline world. Unlike platforms, which observe behavioral fragments within their own boundaries, users directly experience the full continuity of their own lives: their activities across services, their offline routines, their shifting goals, and the private motivations behind observable behavior. This gives users a form of knowledge that no data pipeline can replicate. A user knows not only what they searched for on Google and what they purchased on Amazon, but also why: whether a laptop search was for work or a gift, whether a sudden interest in cooking reflects a new diet or a partner's birthday. In other words, the user can integrate both the implicit behavioral signals that platforms observe and the explicit first-person knowledge that no platform can access. This combination directly addresses the epistemic barrier.

\textbf{Users already have the legal right and practical means to aggregate their data.} Recent data protection frameworks have granted users the legal right to aggregate their own cross-platform data. The EU General Data Protection Regulation (GDPR) Article 20 establishes the right to data portability, requiring platforms to provide users with their personal data "in a structured, commonly used and machine-readable format" and to allow transmission of that data to another controller \citep{eu2016gdpr,protection2018general}. The EU Data Act (Regulation (EU) 2023/2854), applicable from September 2025, extends portability rights beyond personal data to include data generated by connected devices and IoT services \citep{eu2023dataact}. These rights have concrete implementations: Google Takeout allows users to export search history, YouTube activity, email metadata, location history, and more; Amazon provides order history and search history; Apple, Twitter/X, and Meta offer similar export tools.

\textbf{LLM agents as the enabling technology.} The user's data advantage has existed in principle since the early days of data portability rights. In practice, it remained latent: downloading one's data from Google Takeout or Amazon yields gigabytes of JSON, CSV, and HTML files that no ordinary user can interpret, let alone integrate and act upon. LLM agents fundamentally change this equation. LLM agents can reason directly over heterogeneous data formats without requiring a unified schema or any training. They can interpret natural language instructions from users, integrating first-person knowledge ("I'm looking for a new hobby, something hands-on") with structured behavioral data (purchase histories, search logs, watch histories). And they can take actions on the user's behalf: searching for products, filtering recommendations, or interacting with platform APIs.

The agent ecosystem has matured rapidly. Anthropic's Claude Code provides a CLI agent that reads local files, executes code, and interacts with external services \citep{anthropicClaudeCode}. OpenAI's Codex operates as a cloud-based coding agent with sandboxed execution \citep{openai2025codex}. Computer use capabilities allow LLM agents to interact directly with platform interfaces, and the Model Context Protocol (MCP) provides a standardized way for agents to connect to external data sources and tools. These are not research prototypes; they are production systems used daily by millions of users.

Critically, the asymmetry between users and platforms does not lie in reasoning capability. The same LLMs are available to both. Platforms can (and do) deploy frontier models for their recommendation systems. The asymmetry lies in data access: only users can aggregate data across platform boundaries. Given the same LLM, the entity with more complete data about the user will produce better personalization. This is the core structural argument: user-governed personalization is not a normative claim about who \textit{should} control personalization, but an empirical claim about who \textit{can} achieve more comprehensive personalization.

\vspace{-0.5em}
\subsection{Proof-of-Concept Evaluation}
\vspace{-0.5em}

We conducted a proof-of-concept experiment with 15 participants to evaluate whether access to cross-platform user data, combined with LLM-based agents, enables improved personalization.

\textbf{Setup.} Each participant downloaded their personal data exports from multiple platforms, including Amazon (order history, cart activity, digital purchases, and search history), Google Takeout (search activity, shopping activity, and YouTube watch and search history, and more), and, Twitter/X (tweets and likes). All data remained on participants' local machines throughout the study; no raw personal data was shared between participants or uploaded to any central server. Instead, participants reported only aggregate evaluation metrics and summary statistics, without exposing any identifiable personal information. Prior to being processed by the LLM agents, the data was transformed from raw exports into de-identified representations to remove or obfuscate sensitive information. Participants used LLM-based agents built on Claude Code \citep{anthropicClaudeCode}, with models including Sonnet 4.6, Opus 4.6, and Opus 4.7, to process their data and perform personalization tasks. 
\vspace{-0.5em}
\subsubsection{Task 1: Amazon Future-Purchase Prediction}
\vspace{-0.5em}
Given a user's historical behavioral data, the agent is tasked with ranking a candidate set of items that includes both the user's actual future purchases (positives) and 40 negative items sampled from related product categories in the Amazon Review dataset \citep{hou2024bridging}. Each evaluation instance is constructed around a fixed anchor time, with a sliding window where anchor times advance in three-month increments. For each anchor time, the input consists of user activities in the preceding one-year window. The prediction target is the set of items the user actually purchased in the subsequent three-month window. We evaluate ranking quality with standard metrics: hit@$k$, NDCG@$k$, and recall@$k$. Across all participants, this procedure yields a total of 351 evaluation samples. We compare two data conditinos:

\begin{itemize}
\vspace{-0.5em}
\setlength\itemsep{0em}
\item \textbf{Amazon-only}: input data includes Amazon order history, cart activity, digital purchases, and search history within the one-year window preceding the anchor time.
\item \textbf{Amazon + Google cross-platform}: in addition to the Amazon data above, input includes Google Search activity, Google Shopping activity, YouTube watch history, and YouTube search history. Google data is restricted to the window $[\text{anchor\_time} - 1\ \text{year}, \text{anchor\_time} - 7\ \text{days}]$, leaving a one-week buffer to prevent information leakage.
\end{itemize}

\begin{wraptable}{r}{0.5\textwidth}
\vspace{-1em}
\centering
\small
\caption{Amazon future-purchase prediction results.}
\label{tab:amazon}
\resizebox{\linewidth}{!}{
\begin{tabular}{lccc}
\toprule
\textbf{Condition} & \textbf{Hit@5} & \textbf{NDCG@5} & \textbf{Recall@5} \\
\midrule
Amazon-only & 86.6 & 64.8 & 60.1 \\
+ Cross-platform & \textbf{90.0} & \textbf{68.4} & \textbf{63.9} \\
\bottomrule
\end{tabular}
}
\vspace{-1em}
\end{wraptable}

\textbf{Results.} Table~\ref{tab:amazon} summarizes the results. Adding cross-platform Google data improves ranking quality across all metrics. All differences are statistically significant (paired $t$-test, $p < 0.005$ for all three metrics). Cross-platform behavioral data from Google provides useful signals for predicting Amazon purchases, demonstrating that user activity on one platform carries information relevant to personalization on another.

\vspace{-0.5em}
\subsubsection{Task 2: YouTube Video Recommendation.}
\vspace{-0.5em}
The agent generates 20 video recommendations for each participant: 10 \textit{reinforcement} recommendations (grounded in the user's observed YouTube viewing patterns) and 10 \textit{exploration} recommendations (speculative suggestions targeting interests the user may have but has not expressed on YouTube). The agent generates search queries, retrieves real YouTube videos via the YouTube API, and selects one video per query. Participants then watch each recommended video and provide a binary judgment (yes: would want to watch / no: would not). All three conditions are evaluated in a single randomized, blinded session: participants cannot tell which condition produced which recommendation. We compare three conditions:

\begin{itemize}
\vspace{-0.5em}
\item \textbf{YouTube-only}: the agent receives only the participant's YouTube watch history and search history.
\item \textbf{YouTube + full cross-platform}: in addition to YouTube data, the agent receives Google Search activity, Amazon order and search history, and Twitter/X tweets and likes.
\end{itemize}

\begin{wraptable}{r}{0.5\textwidth}
\vspace{-1em}
\centering
\small
\caption{YouTube video recommendation precision.}
\label{tab:youtube}

\begin{tabular}{lccc}
\toprule
\textbf{Condition} & \textbf{Overall} & \textbf{Reinf.} & \textbf{Expl.} \\
\midrule
YouTube-only & 53.3 & 61.5 & 45.3 \\
+ Cross-platform & \textbf{61.6} & \textbf{64.6} & \textbf{58.3} \\
\bottomrule
\end{tabular}

\vspace{-1em}
\end{wraptable}

\textbf{Results.} Table~\ref{tab:youtube} summarizes the results. Cross-platform data improves overall recommendation precision. The most striking pattern is in exploration recommendations: exploration precision, which measures the agent's ability to discover interests the user has \textit{not} expressed on YouTube, improves from 45.3 to 58.3 (+13.0pp). The results show that cross-platform data does not simply reinforce what the platform already knows; it reveals dimensions of the user's interests that are invisible within any single platform.

\paragraph{Summary.} Through the above proof-of-concept experiments, we demonstrate the feasibility of user-governed personalization. These results are preliminary, and fully realizing this potential will require advances which we discuss in Section~\ref{sec:agenda}.
\vspace{-0.5em}
\section{Alternative Views}
\label{sec:alternative}
\vspace{-0.5em}

\subsection{Counterargument 1: Platforms can achieve cross-platform personalization without users.}

\textbf{Federated learning across platforms.} Federated learning (FL) allows multiple parties to jointly train models without sharing raw data, which could in principle enable cross-platform recommendation without requiring users to aggregate their own data. Within a single organization, FL has been deployed successfully: Google trains language models for Gboard across millions of devices using FL with differential privacy \citep{xu2023federated}. One might argue that extending FL across competing platforms could solve the data barrier described in Section \ref{sec:data_barrier}.

\textbf{Ecosystem consolidation and platform-side agents.} Large technology companies already span multiple services. Google operates Search, YouTube, Gmail, Maps, and Photos; Apple spans Messages, Mail, Calendar, and Photos. Both have launched cross-service AI capabilities: Apple Intelligence reasons across on-device data from multiple Apple services \citep{apple2024intelligence}, and Google allows users to connect personal data from multiple Google products to Gemini \citep{pichai2025googleio}. One might argue that continued ecosystem consolidation, combined with platform-side LLM agents, will eventually approximate the cross-platform integration we attribute to users.

\textit{\textbf{Our response.}} {Cross-organization FL for recommendation has not been deployed at scale. A study interviewing 21 practitioners with cross-silo FL experience found that research priorities in federated learning are misaligned with the practical challenges of cross-organization deployment; as one platform practitioner summarized, ``we don't yet see many large-scale success stories of cross-silo FL'' \citep{kuo2025research}. The barriers are organizational rather than technical: misaligned business incentives, trust establishment, and IP concerns. Formal analysis confirms this: market leaders rationally defect from FL coalitions under competitive pressure, because sharing model improvements with competitors erodes their data advantage \citep{meng2024federated}. Existing cross-silo FL deployments cluster in healthcare and pharma, where pre-established legal relationships facilitate cooperation, not among competing consumer platforms. }

{As for ecosystem consolidation, the regulatory constraints described in Section \ref{sec:data_barrier} apply directly: the DMA restricts gatekeepers from combining user data across their own services. More fundamentally, even the most expansive platform ecosystem (Google or Apple) cannot observe a user's activity on competing services (Netflix, Spotify, Amazon) or the user's offline life. Cross-domain recommendation research has documented this structural limitation: user preferences are distributed across platforms in ways that no single ecosystem can capture \citep{zhu2021cross, cantador2015cross}. Ecosystem consolidation narrows the data barrier within one company's products; it does not close it.}

\vspace{-0.5em}
\subsection{Counterargument 2: Platforms have inherent advantages from population-scale collaborative signals.}
\vspace{-0.5em}

\textbf{Population-level behavioral patterns.} Collaborative filtering (CF) \citep{goldberg1992using, koren2009matrix} exploits behavioral patterns across millions of users: "users who bought X also bought Y." This population-level signal is unavailable to a single user, who has only their own data. Platform recommendation systems process billions of interactions to discover latent preference patterns that no individual could observe \citep{covington2016deep,naumov2019deep}. One might argue that this advantage is so large that it cannot be matched by cross-platform personal data, regardless of its completeness.

\textbf{Cold-start and item coverage.} For new items or unfamiliar domains, CF provides a crucial bridge: even when a specific user has no history with a product category, the platform can leverage millions of other users' interactions to make informed recommendations \citep{linden2003amazon}. A user-governed approach, lacking this population signal, might struggle precisely where recommendations are most needed.

\textit{\textbf{Our response.}} {We do not claim that cross-platform personal data replaces collaborative filtering; rather, we claim it addresses a different and complementary dimension. CF answers ``what do users with similar behavior patterns prefer?''; cross-platform personal data answers "what does this specific user's behavior across different services reveal about their interests?" These are orthogonal sources of information. A user who searches Google for marathon training plans, purchases running shoes on Amazon, and listens to running playlists on Spotify is likely interested in fitness-related YouTube content, but no single-platform CF system can detect this cross-domain pattern.}

{Moreover, LLMs partially bridge the cold-start gap that CF traditionally addresses. Recent work demonstrates that LLMs can perform competitively with CF models, because pre-training on large text corpora encodes extensive world knowledge \citep{hou2024large,lin2025recr}. Empirical analysis further shows a possible homomorphism between language representation space and effective recommendation space: item representations derived from language model embeddings can yield strong recommendation performance, suggesting that collaborative-like signals are implicitly encoded within pre-trained language models \citep{sheng2025language}. This does not make CF redundant; integrating explicit collaborative embeddings with LLMs outperforms either approach alone \citep{zhang2025collm}. But it does mean that LLM-based user-governed personalization is not starting from zero: the LLM brings substantial world knowledge that partially compensates for the absence of a user-item interaction matrix. This likely combines population-level collaborative signals with cross-platform personal data, but only the user can provide the latter.
}
\subsection{Counterargument 3: User-governed personalization faces insurmountable practical barriers.}

\textbf{User effort and adoption.} Downloading personal data exports is cumbersome. Google Takeout, Amazon data export, and similar tools require multiple steps, produce large files in heterogeneous formats (JSON, HTML, CSV), and offer no guidance on how to use the exported data. Most users will not invest this effort, especially for an uncertain payoff. If user-governed personalization requires technical sophistication, it will remain a niche capability rather than a mainstream alternative to platform personalization.

\textbf{Data concentration at AI providers.} User-governed personalization, as currently practiced, requires sending personal data to cloud-based LLM providers such as OpenAI or Anthropic. This risks replacing one form of data concentration with another: instead of each platform observing a fragment of the user's behavior, a single AI provider observes the user's entire aggregated digital footprint. This concentration could be worse than the status quo, since the AI provider would hold a more complete picture of the user than any individual platform ever did.

\textbf{Cost and adoption friction.} Running LLM inference over personal data incurs computational costs. Cloud-based APIs charge per token, and processing extensive personal histories (thousands of search queries, years of purchase records) can be expensive. Combined with the effort of downloading data exports, the total cost of user-governed personalization may exceed what most users are willing to bear.

\textit{\textbf{Our response.}} These are genuine obstacles, but they are obstacles of tooling and infrastructure, not of fundamental architecture. The critical distinction is between barriers that are \textit{temporary} and barriers that are \textit{permanent}. The practical barriers to user-governed personalization (export friction, cost, data routing) are tractable engineering problems that are actively being addressed. The data barriers facing platform personalization (competitive incentives, regulatory constraints, epistemic limits) are persistent and not amenable to engineering solutions.

As for data concentration, the current reliance on cloud-based LLM providers is a transitional state, not a necessary feature of user-governed personalization. Local inference hardware is advancing rapidly \citep{nvidia2025dgx}, and the trajectory is clear: models that required cloud infrastructure a year ago already run on consumer devices today. This trend simultaneously mitigates the cost concern, as local inference eliminates per-query API charges. For workloads that still exceed on-device capacity, confidential computing architectures provide a middle path: computation runs on server hardware within trusted execution environments, with cryptographic guarantees that user data cannot be retained or accessed by the service operator \citep{narayan2025proof}. Furthermore, AI providers are subject to the same data protection regulations (GDPR, CPRA) that constrain platforms, and users retain the ability to choose providers, switch between them, or run open-source models entirely locally. The long-term architecture of user-governed personalization is local-first, with cloud inference as an optional fallback.

On adoption friction, the history of personal computing suggests that initially cumbersome capabilities become seamless as tools mature: syncing photos across devices once required manual cable transfers and is now automatic; managing personal finances once required spreadsheets and is now handled by apps that aggregate bank data via APIs. Data export and agent-based personalization are at an early stage of this same trajectory.
\vspace{-0.5em}
\section{Open Problems and Future Directions}
\vspace{-0.5em}
\label{sec:agenda}

Realizing user-governed personalization beyond the proof-of-concept stage requires progress along three fronts: establishing rigorous evaluation methodology, advancing the core modeling capabilities, and building infrastructure for private deployment at scale. We discuss open problems along each front. These fronts are logically sequential: we cannot improve what we cannot measure, and we cannot deploy what we cannot model.

\subsection{Evaluation Methodology}
Our proof-of-concept evaluation in Section \ref{sec:proposal} involves 15 participants and two task families. Scaling this evaluation is a first-order open problem, but the difficulty runs deeper than sample size. Cross-platform behavioral data is inherently private: unlike movie ratings or product reviews, a user's combined search history, purchase records, and media consumption cannot be released as a shared dataset. This rules out the standard paradigm of publishing a static benchmark that the community iterates on.

A more fundamental challenge is methodological. Personalization quality is subjective and multidimensional: a recommendation may be relevant but obvious, surprising but unwanted, or valuable in ways the user cannot articulate in advance. There is no single metric that captures what it means for a recommendation to be "good," and the most informative signal, direct human judgment, is expensive and difficult to scale. Open questions include whether LLM-based evaluation can serve as a reliable proxy for human preference judgments, how to design evaluation protocols that accommodate private data without requiring its release, and what experimental designs can isolate the marginal contribution of cross-platform information from confounding factors such as data volume or recency. Without reliable evaluation methodology, advances in modeling and infrastructure cannot be validated.

\subsection{LLM Agents for Personalization}

\textbf{Training-free approaches versus personalization-aware training.} Even if cross-platform data demonstrably helps, it remains unclear how best to exploit it. Our proof-of-concept relies on off-the-shelf LLMs, coordinated through the agentic infrastructure of Claude Code, to process raw data exports without any model adaptation. Whether this represents a floor or a ceiling is an open question. At one end of the spectrum, advances in prompt design and context engineering may suffice: the LLM's pre-trained world knowledge, combined with well-structured user data and a well-designed agentic framework, may already capture the relevant preference signals. At the other end, personalization may require fundamentally different training objectives. Current LLMs are aligned on general-purpose objectives, e.g., correctness in code and mathematics, helpfulness in dialogue, and harmlessness in safety-critical contexts. None of these directly optimize for understanding individual human preferences, which are often inconsistent, context-dependent, and not fully rational \citep{kleinberg2024challenge}. When a model or an agent has access to a user's complete cross-platform behavioral history, the question of what training objective best serves personalization becomes a first-class research problem.

\textbf{Recovering collaborative signals without centralization.} A related limitation is the absence of population-level collaborative signals. Section \ref{sec:alternative} acknowledged that user-governed personalization cannot access the "users who bought X also bought Y" patterns that platforms observe across millions of users. A natural direction is to develop federated protocols in which users voluntarily share anonymized preference representations (not raw data) to benefit from collective intelligence while retaining data sovereignty. Unlike platform-side federated learning, where the aggregation server is controlled by a platform with its own commercial incentives, user-side federation would be structured so that users are the data owners and participation is voluntary. Key open problems include designing privacy-preserving aggregation mechanisms (secure aggregation, differential privacy) that work with heterogeneous, cross-platform user representations, and understanding what granularity of shared information maximizes collective benefit while minimizing privacy loss.

\subsection{Infrastructure for Private Deployment}

As argued in Section \ref{sec:alternative}, user-governed personalization should not depend on proprietary cloud APIs that create new data concentration risks. The open-source community should prioritize models optimized for personalization workloads: long-context reasoning over heterogeneous personal data, preference modeling, and recommendation generation. Models such as Qwen \citep{yang2025qwen3} provide a foundation, but personalization-specific capabilities remain underdeveloped. Edge-side inference must also advance to meet the demands of personal data processing: reasoning over years of cross-platform behavioral data requires progress in context compression, retrieval-augmented generation over personal data stores, and efficient long-context techniques.

\section{Conclusion}
This paper argues for a shift from platform-centric personalization to user-governed personalization. The user is the only entity that spans all platforms and the offline world; data portability rights provide legal access to cross-platform data, and LLM agents now make it practically actionable. Realizing user-governed personalization at scale requires progress in evaluation methodology, personalization-aware modeling, and privacy-preserving infrastructure. The window to shape how personal AI develops is open now: researchers, developers, and policymakers should prioritize building the tools, models, and safeguards that keep users in control of their own personalization.

\newpage

\bibliographystyle{plain}
\bibliography{neurips_2026}

\newpage

\appendix
\section{Ethics Statement}
\label{app:ethics}

This study involved 15 human participants who voluntarily provided their personal data exports for evaluating user-governed personalization. All participants gave informed consent before participating. The study protocol was designed to minimize privacy risks:

\begin{itemize}
\item All personal data remained on each participant's own local machine throughout the study. Researchers never accessed, collected, or stored any participant's personal data.
\item Prior to being processed by LLM agents, participants' data was transformed from raw platform exports into de-identified representations to remove or obfuscate sensitive information (e.g., names, email addresses, and other personally identifiable information).
\item Each participant independently ran the evaluation pipeline on their own machine and submitted only evaluation metrics (e.g., hit rates, NDCG scores, precision values) and summary statistics. No identifiable personal information was shared with the research team.
\item LLM Agent calls were made from each participant's own machine using their own Claude Code accounts.
\end{itemize}

All participants are affiliated with the research team and participated on a voluntary basis. No external recruitment or compensation was involved.

\section{Declaration of LLM Usage}
\label{app:llm_usage}

We use large language models (LLMs) in the following ways:

\textbf{Paper writing.} LLMs were used to assist with polishing this paper. The intellectual contributions, arguments, experimental design, and analysis are entirely the work of the human authors.

\textbf{Proof-of-concept experiments.} The proof-of-concept evaluation described in Section~\ref{sec:proposal} used Claude Code \citep{anthropicClaudeCode} as the core experimental infrastructure. Participants used LLM-based agents built on Claude Code, with models including Sonnet 4.6, Opus 4.6, and Opus 4.7, to process their personal data exports and perform the personalization tasks (Amazon future-purchase prediction and YouTube video recommendation).


\end{document}